\documentclass[aps,prb,twocolumn,superscriptaddress]{revtex4-2}
\usepackage{graphicx}
\usepackage{latexsym}
\usepackage{amssymb}
\usepackage{amsmath}
\usepackage{amsfonts}
\usepackage[vcentermath]{youngtab}
\usepackage{upgreek}
\usepackage{bm,dsfont}
\usepackage{multirow}
\usepackage{enumitem}
\usepackage{color}
\usepackage{hyperref}
\hypersetup{
colorlinks = true,
linkcolor = [rgb]{0.70,0.13,0.13},
citecolor = [rgb]{0.13,0.55,0.13},
urlcolor = [rgb]{0.25, 0.41, 0.88}}
\usepackage{makecell}

\newcommand{\bra}[1]{\langle #1|}
\newcommand{\ket}[1]{|#1 \rangle}

\newcommand{\dd}{\mathrm{d}}

\newcommand{\ii}{\mathrm{i}}
\newcommand{\e}{\mathrm{e}}

\newcommand{\U}{\mathrm{U}}
\newcommand{\SU}{\mathrm{SU}}

\newcommand{\spin}{\mathfrak{spin}}

\newcommand{\dsZ}{\mathbb{Z}}

\newcommand{\scL}{\mathcal{L}}

\newcommand{\vect}[1]{{\bm{#1}}}

\newcommand{\eqnref}[1]{Eq.\,\eqref{#1}}
\newcommand{\figref}[1]{Fig.\,\ref{#1}}
\newcommand{\tabref}[1]{Tab.\,\ref{#1}}

\newcommand{\appref}[1]{Appendix\,\ref{#1}}
\newcommand{\refcite}[1]{Ref.\,\onlinecite{#1}}

\def \Z{\mathbb{Z}}

\newcommand{\bea}{\begin{eqnarray}}
\newcommand{\eea}{\end{eqnarray}}
\def\be{\begin{equation}}
\def\ee{\end{equation}}
\newcommand{\beq}{\begin{equation}}
\newcommand{\eeq}{\end{equation}}
\newcommand{\beqn}{\begin{eqnarray}}
\newcommand{\eeqn}{\end{eqnarray}}

\begin{document}

\title{Variational Monte Carlo Study of Symmetric Mass Generation in a Bilayer Honeycomb Lattice Model}

\author{Wanda Hou}
\author{Yi-Zhuang You}
\affiliation{{Department of Physics, University of California at San Diego, La Jolla, CA 92093, USA}}
\date{\today}

\begin{abstract}
We investigate a bilayer honeycomb lattice model of spin-1/2 fermions at half-filling with local Heisenberg coupling of fermion spins across the two layers. Using variational Monte Carlo (VMC) simulation, we demonstrate that the system undergoes a direct transition from a Dirac semimetal phase to a symmetric gapped phase, known as symmetric mass generation (SMG), as the Heisenberg coupling strength is increased. The transition does not involve spontaneous symmetry breaking or topological order and has been proposed as an example of the fermionic deconfined quantum critical point (fDQCP). Our simulation shows that a fermionic parton bilinear mass opens at the transition point while all symmetries are still preserved thanks to the quantum fluctuations introduced by the correlation factor in the variational wave function. From the simulation data, we extract the critical exponent $\nu=0.96\pm0.03$ and the fermion scaling dimension $\Delta_c=1.31\pm0.04$ at the SMG critical point, which are consistent with the field theoretical prediction of fDQCP in (2+1)D. These findings support the hypothesis that the fermion fractionalizes at the SMG critical point.
\end{abstract}

\maketitle

\section{Introduction}

Symmetric mass generation (SMG) \cite{Wang2204.14271} is a mechanism by which gapless Dirac fermions acquire a gap in their excitation spectrum without breaking symmetry or developing topological order. The gap-opening transition, known as the SMG transition, involves non-perturbative interaction effects among the fermions and does not require any fermion bilinear mass condensation, in contrast to the conventional Higgs mechanism \cite{Anderson1963pcPRPlasmons, EnglertBrout1964PRL, Higgs1964PRL, Guralnik1964PRL13.585, Weinberg1967}. Numerical simulations \cite{Slagle:2015lo, Ayyar2015Massive, Catterall:2016sw, Ayyar:2016fi, He:2016qy, Ayyar2017Generating, Butt2018SO4-invariant, Catterall2020Exotic, Butt2021Symmetric, Zeng2022Symmetric, Hasenfratz2204.04801} have shown that the SMG transition can be a direct and continuous quantum critical point in various models, which is intriguing because it does not fit the Landau-Ginzburg-Wilson paradigm, as the phases on either side are not distinguished by any symmetry-breaking order parameter. It has been suggested \cite{You2018Symmetric, You2018From} that the SMG critical point is described by a \emph{fermionic} version of the deconfined quantum critical point (DQCP) \cite{Senthil:2004wj, Motrunich:2004hh, Senthil:2004qm, Senthil:2006lz}, in which fermions \emph{fractionalize} into deconfined bosonic and fermionic partons at (and only at) the critical point. However, numerical evidence for the fermion fractionalization is still lacking.

The goal of this study is to understand the universal properties of the SMG transition and provide evidence for the fermion fractionalization hypothesis. To this end, we examine a bilayer honeycomb lattice model of interacting fermions and investigate the SMG transition using numerical and analytical methods. On the numerical side, we propose a variational ansatz for the quantum many-body ground state of the fermion system and use the variational Monte Carlo (VMC) approach to identify and simulate the SMG transition. On the theoretical side, we develop a quantum field theory description of the SMG critical point as a fermionic DQCP (fDQCP) and predict its universal properties using renormalization group (RG) analysis. Our numerical results are in good agreement with the analytic predictions, supporting the fractionalization of fermions at the SMG transition.

\section{Lattice Model} 

\begin{figure}[htbp]
\begin{center}
\includegraphics[width=180pt]{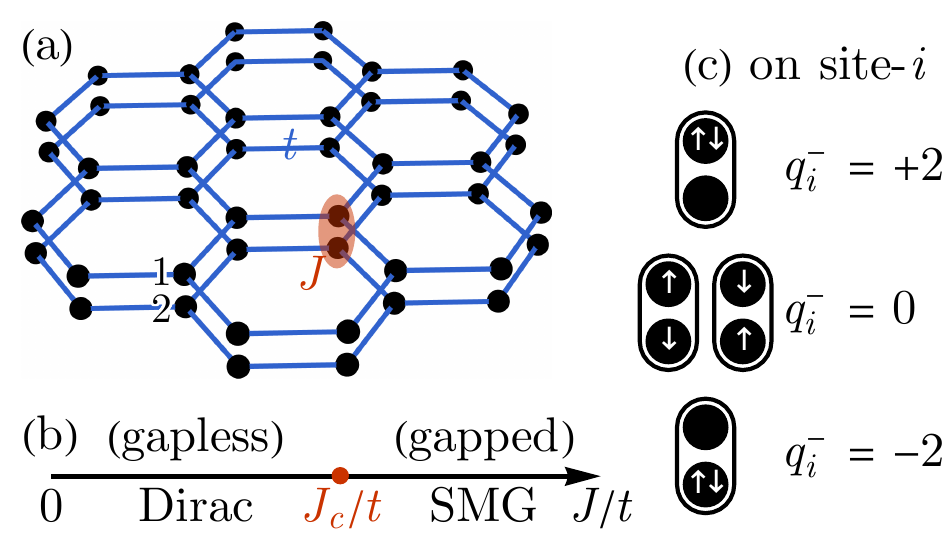}
\caption{(a) The bilayer honeycomb lattice model with intra-layer hopping $t$ and inter-layer Heisenberg coupling $J$. (b) The phase diagram as $J/t$ varies. (c) On-site fermion configurations of different inter-layer $\U(1)^{-}$ charges $q_i^-$.}
\label{fig: lattice}
\end{center}
\end{figure}

We investigated a (2+1)D example of SMG. The model is defined on an A-A stacking bilayer honeycomb lattice, as illustrated in \figref{fig: lattice}(a), with each site $i$ hosting four complex fermion modes, denoted as $c_{il\sigma}$, where $l=1,2$ labels the two layers and $\sigma=\uparrow, \downarrow$ labels the two spin components. The model is described by the following Hamiltonian
\begin{equation}\label{eq: H}
    H=-t\sum_{\langle ij \rangle,l,\sigma}(c^{\dagger}_{il\sigma}c_{jl\sigma}+\text{h.c.})+J\sum_{i}\vect{S}_{i1}\cdot\vect {S}_{i2},
\end{equation}
where $\langle ij \rangle$ stands for the nearest neighboring pairs of sites $i$ and $j$ on the honeycomb lattice. There is no interlayer fermion hopping. The two layers are only coupled by the antiferromagnetic Heisenberg interaction $J>0$. $\bm{S}_{il}=\frac{1}{2}c^{\dagger}_{il\alpha}\vect{\sigma}^{\alpha\beta}c_{il\beta}$ denotes the spin operator at site-$i$ layer-$l$, with $\vect{\sigma}=(\sigma_x,\sigma_y,\sigma_z)$ being the Pauli matrices. The relative interaction strength $J/t$ is the only tuning parameter of this model. 

Consider the fermion system at half-filling (with zero chemical potential). When $J/t=0$, the model describes two decoupled layers of graphene, which produces eight gapless Dirac fermions (two valleys $\times$ two layers $\times$ two spins) at low energy. The gapless \emph{Dirac semimetal} phase is stable against small $J/t$ perturbation, as the local interaction $J$ is irrelevant for (2+1)D Dirac fermions under RG. When $J/t\to\infty$, the inter-layer Heisenberg interaction acts on each honeycomb site independently, which leads to a unique gapped ground state $\bigotimes_{i}(c^{\dagger}_{i1\uparrow}c^{\dagger}_{i2\downarrow}-c^{\dagger}_{i1\downarrow}c^{\dagger}_{i2\uparrow})\ket{0}$ as the direct product of inter-layer spin-singlet states. Such an insulating state will be called an \textit{SMG insulator}, as it does not break any symmetry (either lattice or internal) of the model. Given that any band insulator in this half-filled bilayer honeycomb lattice model would necessarily break some symmetry (see \appref{app: sym} for symmetry analysis), the interaction effect is essential in achieving the SMG insulator. Therefore, as $J/t$ is increased, we expect a phase diagram as shown in \figref{fig: lattice}(b), where an \emph{SMG transition} happens at some critical $J_c$ (to be determined), separating the Dirac semimetal and the SMG insulator. The model could be relevant to the {twisted bilayer graphene} \cite{Cao1802.00553, Cao1803.02342}, which has the same Dirac fermion content at low energy, although the interaction may be different. Nevertheless, investigating this toy model is instructive for the future experimental realization of SMG physics in correlated materials.

\section{VMC Simulation}

\subsection{Method}

The (2+1)D SMG transition has been numerically observed \cite{Slagle:2015lo,He:2016qy} in similar models by quantum Monte Carlo (QMC) simulations. The results suggest that the SMG transition is direct and continuous, but the critical exponent $\nu$ (of the correlation length $\xi\sim|J-J_c|^{-\nu}$ scaling) and the fermion operator scaling dimension were not determined yet. To investigate these universal properties of the SMG transition, we proposed a variational Monte Carlo (VMC) approach  \cite{Ceperley1977VMC, Sorella0009149}, based on a variational wave function motivated by the parton-Higgs theory \cite{You2018Symmetric}. \refcite{You2018Symmetric} proposes a non-Abelian gauge theory that involves a more complicated confinement mechanism to enter the SMG phase. This work simplifies the field theory, reducing the gauge group to the Abelian group, which allows the direct application of known RG analysis.

Compared to the QMC approach, although the VMC approach is biased by the variational ansatz, it can be pushed to relatively larger system sizes with fewer computational resources. Moreover, the variational wave function provides a unique insight that was not available by other methods, i.e.,~the mean-field mass. This approach allows us to effectively observe the gap opening of the fermionic parton, which deepens our understanding of the underlying parton physics across the SMG transition.

The construction of the variational state starts with a mean-field (free-fermion) Hamiltonian $H_\text{MF}[\lambda]=-t\sum_{\langle ij\rangle,l}(c_{il}^\dagger c_{jl}+\text{h.c.})+\lambda\sum_{i}(-)^i(c_{i1}^\dagger c_{i2}+\text{h.c.})$ (with $c_{il}=[c_{il\uparrow}, c_{il\downarrow}]^\intercal$ including both spins, and spin indices are summed over in $H_\text{MF}$ implicitly). It introduces an inter-layer hopping term $\lambda$ with a stagger sign $(-)^i=\pm$ between $A/B$ sublattices on the honeycomb lattice. This stagger sign is important to make the $\lambda$ term to be a mass term that gaps out the Dirac fermions on the mean-field level. $\lambda$ will be treated as a variational parameter. Consider the ground state $\ket{\Psi_\text{MF}[\lambda]}$ of $H_\text{MF}[\lambda]$ followed by a soft projection (correlation factor) that depends on the configuration of inter-layer charge difference $q_i^-:=c_{i1}^\dagger c_{i1}-c_{i2}^\dagger c_{i2}$ on every site, as exemplified in \figref{fig: lattice}(c), the variational many-body state then takes the following form, similar to a Slater-Jastrow wave function \cite{Jastrow1955}
\begin{equation}\label{eq: Psi}
\ket{\Psi[\lambda,V]}\propto \e^{-V[\{q_i^-\}]}\ket{\Psi_\text{MF}[\lambda]},
\end{equation}
where the Jastrow-like energy functional $V[\{q_i^-\}]=\sum_i V(q_i^-)$ will be variational optimized as well. The objective is to optimize the varitional state by minimizing the energy expectation value $\bra{\Psi[\lambda,V]}H\ket{\Psi[\lambda,V]}$ given the physical Hamiltonian $H$ in \eqnref{eq: H}. 

The variational state in \eqnref{eq: Psi} is designed to reproduce the exact ground states in both the strong interaction ($J/t\to\infty$) and the weak interaction ($J/t\to 0$) limits. In the strong interaction limit, the extreme SMG state corresponds a product of inter-layer singlets, which can be obtained by taking the limit of strong inter-layer hopping ($\lambda/t\to\infty$) followed by the hard projection $\e^{-V[\{q_i^-\}]}\propto \prod_i\delta_{q_i^-=0}$ that imposes $q_i^-=0$ on every site. On the other hand, the exact free-fermion ground state at $J/t= 0$ is achieved by turning off $\lambda$ and $V$. This design ensures that the variational state accurately captures the behavior of the system in both limits.

For intermediate $J/t$, we evaluated the variational energy $\bra{\Psi[\lambda,V]}H\ket{\Psi[\lambda,V]}$ by the VMC sampling technique (see \appref{app: VMC} for algorithm details) and minimized it using the PyTorch library \cite{PyTorch} with the Adam optimizer \cite{Kingma1412.6980}. This allows us to optimize the variation state by leveraging the auto-differentiation capabilities of PyTorch. To verify the accuracy of our algorithm, we conducted a benchmark test with exact diagonalization in a single unit cell (consisting of two sites), which shows that our variational ground state can achieve a fidelity of at least $0.99$ and has a relative energy excess of at most $10^{-4}$ across all values of $J/t$.

\begin{figure}[htbp]
\begin{center}
\includegraphics[width=246pt]{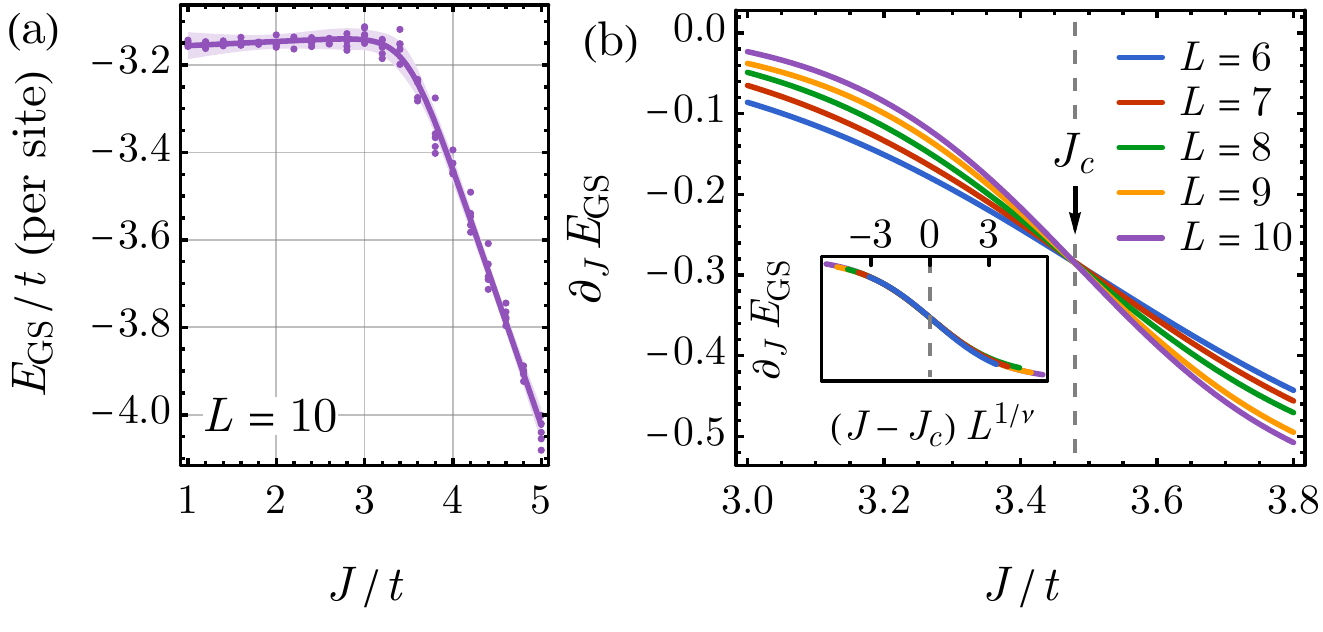}
\caption{(a) Ground state energy $E_\text{GS}$ and (b) its first-order derivative $\partial_J E_\text{GS}$ with respect to the interaction strength $J$. Each dot in (a) represents an estimation of $E_\text{GS}$ from one VMC simulation.}
\label{fig: energy}
\end{center}
\end{figure}

\subsection{Numerical Results}

In our study, we set the energy unit to be $t=1$ and plotted the ground state energy $E_\text{GS}$ as a function of the interaction strength $J$ in \figref{fig: energy}(a). To estimate $E_\text{GS}$ at each value of $J$, we conducted multiple VMC simulations with different initial random seeds and fit a smooth curve to the data. The first-order derivative $\partial_J E_\text{GS}$ was then computed based on the fitted energy curve, as shown in \figref{fig: energy}(b). The calculations were performed on lattices of $L\times L$ unit cells (totally $2L^2$ sites) for $L=6,7,\cdots,10$. We observed that all curves cross at a single point, which we identified as the SMG critical point $J_c=3.48$. By rescaling the horizontal axis to $(J-J_c)L^{1/\nu}$, we were able to collapse all curves onto a single curve when the exponent was tuned to $\nu=0.96\pm0.03$, as shown in the inset of \figref{fig: energy}(b). This determines the power-law scaling of the correlation length $\xi\sim |J-J_c|^{-\nu}$ near the SMG transition.

\begin{figure}[htbp]
\begin{center}
\includegraphics[width=220pt]{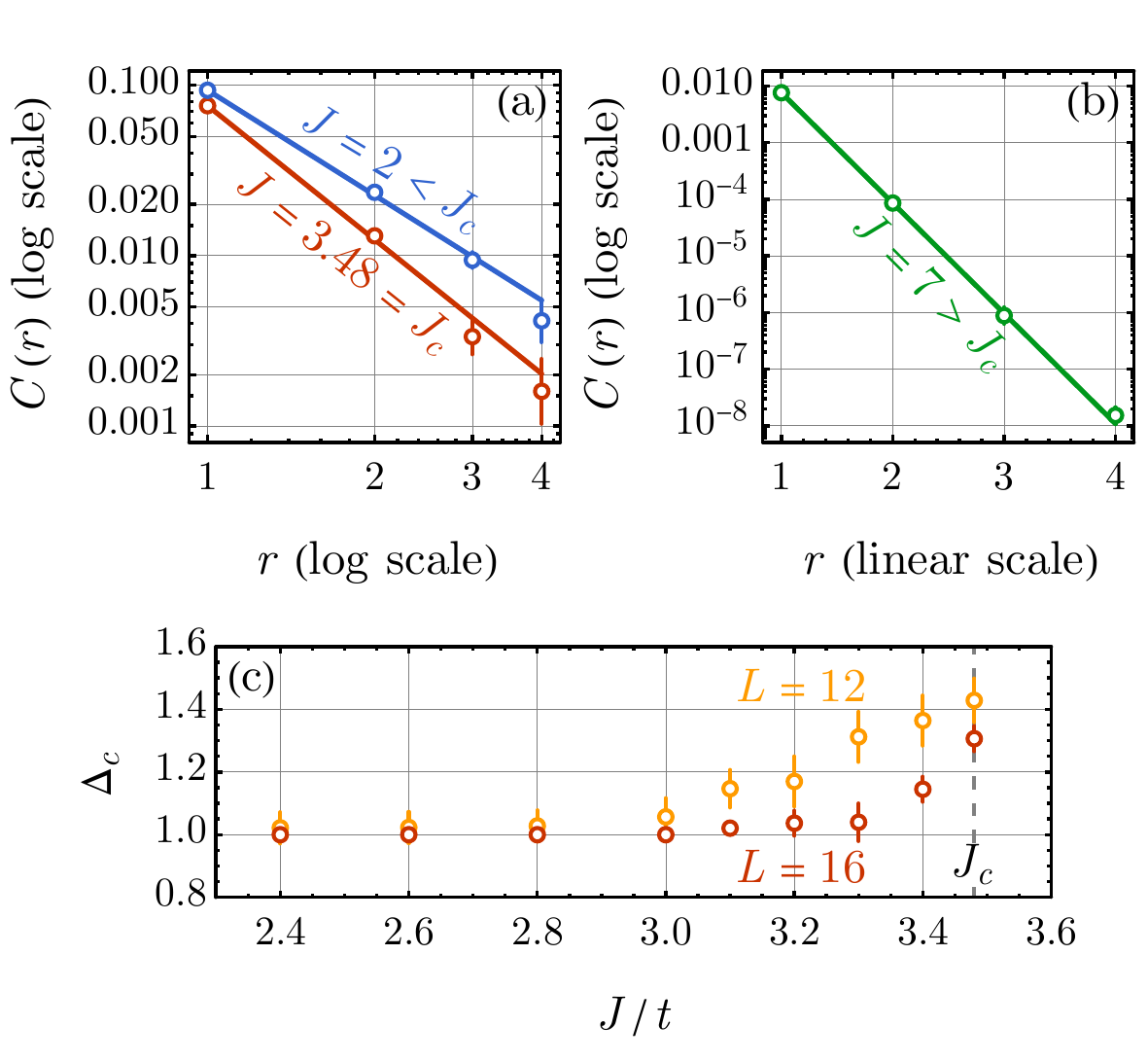}
\caption{Fermion-fermion correlation $C(r)$ for (a) $J\leq J_c$ and (b) $J>J_c$, computed on the $L=16$ lattice. (c) Fermion scaling dimension $\Delta_c$ for $J\leq J_c$.}
\label{fig: correlation}
\end{center}
\end{figure}

To investigate the fermion scaling dimension at the critical point, we measured the fermion-fermion correlation function $C(r):=\bra{\Psi[\lambda,V]} c_{il}^\dagger c_{jl}\ket{\Psi[\lambda,V]}$ between two sites $i$ and $j$ as a function of their separation distance $r=|\vect{r}_i-\vect{r}_j|$ on a lattice of size $L=16$. The correlation function $C(r)$ exhibits a power-law behavior $C(r)=1/r^{2\Delta_c}$ for $J\leq J_c$, as plotted in \figref{fig: correlation}(a) with the log-log scale. The exponent $\Delta_c$ corresponds to the scaling dimension of the fermion operator $c_{il}$. We found that the scaling dimension in the Dirac semimetal phase (at $J=2<J_c$ for example) is $\Delta_c=1.00\pm0.03$, which is the same as that of the free Dirac fermion in (2+1)D, indicating that the local interaction $J$ is perturbatively irrelevant in this phase. However, at the SMG critical point ($J=J_c$), we observed a distinct scaling dimension of $\Delta_c=1.31\pm0.04$, which suggests a different RG fixed point. In the gapped phase (at $J=7>J_c$ for example), we verified that $C(r)$ becomes short-ranged and decays exponentially as expected, as plotted in \figref{fig: correlation}(b) with the log-linear scale. This clearly signifies that the SMG transition is a semimetal-insulator (gapless to gapped) transition.

The enlarged fermion scaling dimension at the SMG critical point indicates the fractionalization of fermions. To validate this observation more systematically, we extracted the fermion scaling dimension $\Delta_c$ by fitting the power-law correlation at different values of $J\leq J_c$ for two system sizes $L=12,16$, as shown in \figref{fig: correlation}(c). As the system size increases (from $L=12$ to $L=16$), we observed the trend that $\Delta_c$ approaches 1 for $J<J_c$ but seems to converge to some finite value above 1 at $J=J_c$. Limited by the available system sizes and the data quality, we were unable to extrapolate our results to infinite system size reliably. Nevertheless, the evidence clearly requires a different conformal field theory (CFT) description of the SMG critical point distinct from the free fermion CFT.

\begin{figure}[htbp]
\begin{center}
\includegraphics[width=240pt]{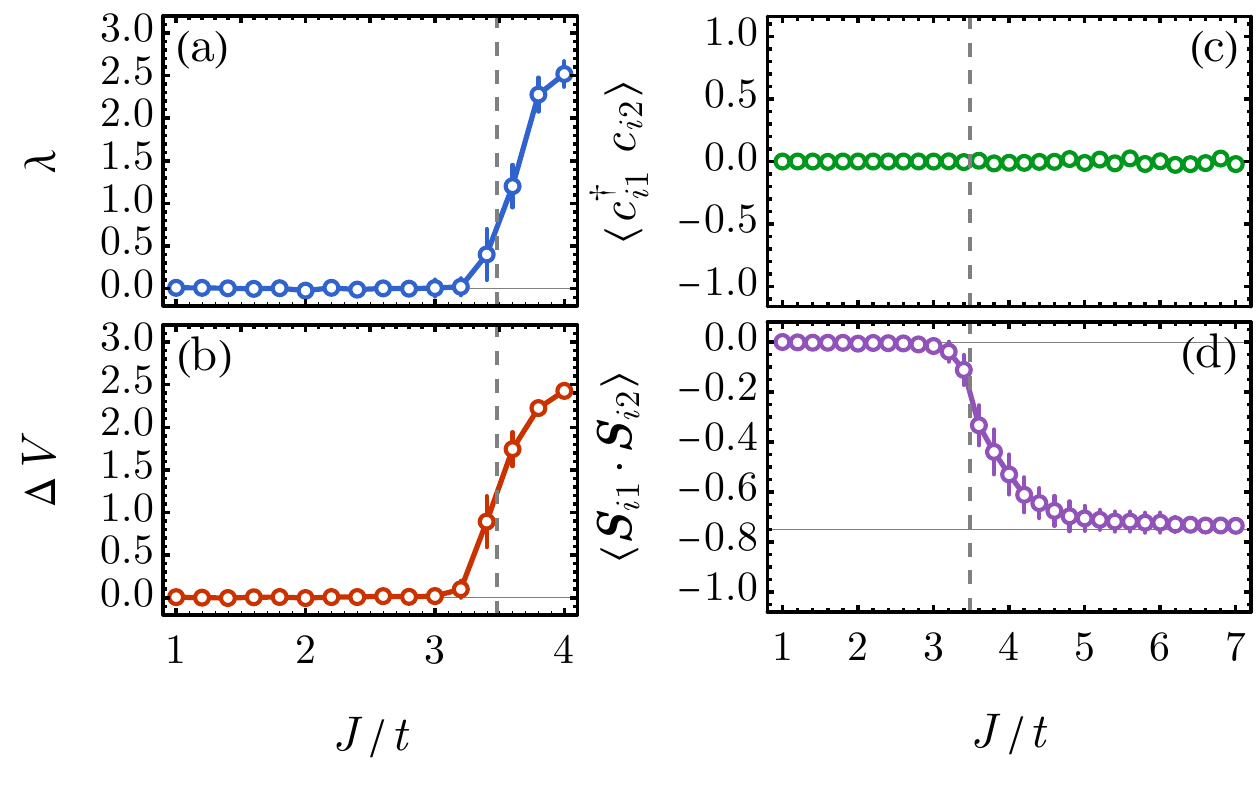}
\caption{(a) The inter-layer hopping strength $\lambda$ in the mean-field Hamiltonian $H_\text{MF}[\lambda]$. (b) The gap of Jastrow energy $\Delta V=V(\pm2)-V(0)$ between $q^-=\pm2$ and $q^-=0$ configurations, characterizing the strength of the soft projector. (c) The inter-layer local fermion coherence (tunneling amplitude). (d) The inter-layer local spin-spin correlation. The results are obtained by finite-size scaling to $L\to\infty$ based on $L=5,\cdots,10$ VMC simulations (see \appref{app: FSS}).}
\label{fig: VMC}
\end{center}
\end{figure}

To better understand the SMG mechanism, we investigated how the variational parameters change across the SMG transition. We found that the staggered inter-layer hopping parameter $\lambda$ in the mean-field Hamiltonian $H_\text{MF}[\lambda]$ turns on continuously near the transition, as in \figref{fig: VMC}(a), resulting in a fermion gap opening in the mean-field state $\ket{\Psi_\text{MF}[\lambda]}$. This gap is sufficient to make the fermion-fermion correlation $C(r)$ short-ranged and generate a mass for all fermion excitations in the SMG insulator phase. However, such kind of mass generation comes with the spontaneous breaking of the layer $\U(1)^{-}$ global symmetry that is associated with the conservation of the total inter-layer charge difference $Q^{-}=\sum_{i}q_i^-$. Under the $\U(1)^{-}$ symmetry, the inter-layer hopping term $c_{i1}^\dagger c_{i2}$ transforms as a charge-2 operator. On the mean-field level, a finite $\lambda$ will induce a finite $\langle c_{i1}^\dagger c_{i2}\rangle$ expectation value on the mean-field state $\ket{\Psi_\text{MF}[\lambda]}$, which breaks the $\U(1)^{-}$ symmetry to its $\dsZ_{2}^{-}$ subgroup.

Fortunately, the correlation factor $\e^{-V[\{q_i^-\}]}$ helps to restore the $\U(1)^{-}$ symmetry without closing the fermion excitation gap. Recall that the Jastrow-like energy function $V[\{q_i^-\}]=\sum_i V(q_i^-)$ was written as a sum of on-site potentials. We found that the local potential difference $\Delta V=V(\pm2)-V(0)$ increases across the SMG transition, as shown in \figref{fig: VMC}(b). The possitive $\Delta V$ favors the $\U(1)^{-}$ neutral ($q_i^-=0$) over the $\U(1)^{-}$ charged ($q_i^{-}=\pm2$) configurations on every site (see \figref{fig: lattice}(c) for examples). As the potential difference $\Delta V$ increases, the correlation factor $\e^{-V[\{q_i^-\}]}$ suppresses local $\U(1)^{-}$ charge fluctuations, effectively introducing $\U(1)^{-}$ gauge (phase) fluctuations \cite{Motrunichcond-mat/0412556, Hermele0803.1150, Weng1105.3027} due to the quantum mechanical uncertainty relation between the charge and phase observables. As a result, the $\langle c_{i1}^\dagger c_{i2}\rangle$ expectation value is disordered by the gauge fluctuation, and only those $\U(1)^{-}$ symmetric observables, such as $\langle \vect{S}_{i1}\cdot \vect{S}_{i2}\rangle$, survive the gauge fluctuation, as clearly shown in \figref{fig: VMC}(c,d). This establishes the $J>J_c$ phase as a \emph{symmetric gapped} phase, where the fermion mass (gap) is generated by the condensation of multi-fermion operators (such as $\vect{S}_{i1}\cdot \vect{S}_{i2}$) instead of two-fermion operators (such as $c_{i1}^\dagger c_{i2}$).

\section{Field Theory Analysis}

The above discussion motivates the fermion fractionalization hypothesis, which states that the physical fermion $c_{il}$ fractionalizes into deconfined bosonic $b_{il}$ and fermionic $f_{il}$ partons at (and only at) the SMG critical point (hence an fDQCP), as
\begin{equation}\label{eq: c=bf}
    c_{il}=\begin{bmatrix}c_{il\uparrow} \\ c_{il\downarrow}\end{bmatrix}=b_{il}^\dagger \begin{bmatrix}f_{il\uparrow} \\ f_{il\downarrow}\end{bmatrix}=b_{il}^\dagger f_{il},
\end{equation}
for each layer $l=1,2$ separately. A parton mean-field analysis can be found in \appref{app: MF}. The fermionic parton $f_{il}$ still carries the $\SU(2)$ spin quantum number and hops on the honeycomb lattice. However, the separate charge conservation symmetry in each layer $l$, denoted as $\U(1)_l$, is now acting on the bosonic parton $b_{il}$ only. Note that the previously mentioned layer $\U(1)^{-}$ symmetry is a subgroup of the larger $\U(1)_1\times\U(1)_2$ symmetry.

The partons $b_{il},f_{il}$ are redundant descriptions of the physical fermion $c_{il}$, because the following transformations have no physical effect on \eqnref{eq: c=bf}
\begin{equation}
\tilde{\U}(1)_l:b_{il}\to \e^{\ii\theta_{il}}b_{il}, f_{il}\to \e^{\ii\theta_{il}}f_{il}.
\end{equation}
They are identified as the emergent gauge group $\tilde{\U}(1)_1\times\tilde{\U}(1)_2$ arising from the fermion fractionalization. The low-energy physics of the SMG transition in the bilayer honeycomb model can then be described by the following field theory in the parton language
\begin{equation}\label{eq: L}
\mathcal{L}=\sum_{l=1,2}|(\partial-\ii (a_l-A_l))b_l|^2+r|b_l|^2 + \bar{f}_l\,\gamma\cdot(\partial-\ii a_l)f_l,
\end{equation}
where $l=1,2$ labels the two layers. Within each layer $l$, the theory describes a single-flavor ($N_b=1$) bosonic scalar field $b_l$ coupled to a four-flavor ($N_f=4$) fermionic spinor field $f_l$ through the $\tilde{\U}(1)_l$ gauge field $a_l$. The four internal flavors of $f_l$ stand for the spin and valley degrees of freedom inherited from the physical fermion $c_l$, and $\gamma$ denotes the $\gamma$ matrices in the Dirac spinor space. $A_l$ is the background gauge field that track the physical $\U(1)_l$ symmetry. 

In this field theory \eqnref{eq: L}, the SMG transition is driven by tuning the only parameter $r$:
\vspace{-3pt}
\begin{itemize}[leftmargin=*]
\setlength\itemsep{-3pt}
\item When $r<0$, the scalar fields $b_l$ condense, pinning the gauge fields $a_l$ to the symmetry background fields $A_l$ through the Higgs mechanism, such that the fermionic partons $f_l$ regain the $\U(1)_1\times\U(1)_2$ symmetry charges and restore the gapless physical fermions in the weakly-interacting Dirac semimetal phase.
\item When $r>0$, the scalar fields $b_l$ are gapped and decoupled from the theory. Driven by the gauge interaction, the fermionic partons can spontaneously develop a parton Higgs mass $\lambda (\bar{f}_1 f_2+\text{h.c.})$, which gaps out all fermions and Higgs the $\tilde{\U}(1)_1\times\tilde{\U}(1)_2$ gauge structure down to the diagonal $\tilde{\U}(1)^{+}$ \footnote{The $\dsZ_2$ subgroup of $\tilde{\U}(1)^{-}$ is shared with $\tilde{\U}(1)^{+}$, so the Higgsing does not lead to $\dsZ_2$ topological order as well.}. Then the diagonal $\tilde{\U}(1)^{+}$ gauge field automatically confines at low energy by the Polyakov mechanism \cite{Polyakov1977}, leading to a trivially-gapped SMG phase. Since $A_l$ does not couple to the parton Higgs field $\lambda$, the $\U(1)_1\times\U(1)_2$ symmetry remains unbroken even if $\lambda$ has condensed. 
\end{itemize}
\vspace{-3pt}
The SMG critical point is therefore described by the field theory \eqnref{eq: L} at $r=0$. The state of various fields across the transition is summarized in \tabref{tab: phases}

\begin{table}[ht]
\begin{tabular}{c|c|c|c}
 \ & Dirac semimetal & SMG transition & SMG insulator\\
 \hline\hline
 \ $c_{l}$ & gapless & fractionalized & gapped \\
 \hline
 \ $b_{l}$ & condensed ($r<0$) & critical ($r=0$) & gapped ($r>0$) \\
 \hline
 \ $f_{l}$ & gapless ($\lambda=0$) & gapless ($\lambda=0$) & gapped ($\lambda\neq 0$) \\
 \hline
 \multirow{2}{*}{\ $a_{l}$} & \multirow{2}{*}{Higgsed} & \multirow{2}{*}{deconfined} & $a^-$ Higgsed \\
 & & & $a^+$ confined \\
\end{tabular}
\caption{State of physical fermion $c_l$, bosonic $b_l$ and fermionic $f_l$ parton, and gauge fields $a_l$ across SMG transition. $a^\pm = a_1\pm a_2$ are linear combinations of $a_l$.}
\label{tab: phases}
\end{table}

To estimate the scaling dimension $\Delta_c$ of the physical fermion $c_l\sim b_l^\dagger f_l$, we extended the quantum electrodynamics (QED) theory to general $N_b$ bosonic flavors and $N_f$ fermionic flavors and used the result of RG analysis \cite{Kaul0801.0723, Benvenuti1902.05767} by the large-$N_b,N_f$ expansion, which predicted
\begin{equation}\label{eq: Delta_c}
\Delta_c=\frac{3}{2}+\frac{2}{3\pi^2N_b}-\frac{40}{3\pi^2(N_b+N_f)}+\cdots.
\end{equation} 
Extrapolating the result to $(N_b, N_f)=(1,4)$, \eqnref{eq: Delta_c} predicts $\Delta_c\simeq 1.3>1$, which is consistent with our VMC simulation. This provides supportive evidence for the fermion fractionalization hypothesis and establishes the SMG transition as an fDQCP.

\section{Summary}

We developed a VMC approach to simulate a bilayer honeycomb lattice model and investigated the critical behavior of the SMG transition in this model. We tested the fermion fractionalization hypothesis by measuring the fermion scaling dimension $\Delta_c= 1.31\pm 0.04$ at the critical point and showing that the result is consistent with the prediction for an fDQCP, $\Delta_c\simeq 1.3>1$. The fDQCP field theory indicates that the mean-field parameter $\lambda$ in the variational state should be interpreted as the Higgs mass of the fermion parton as $\lambda(\bar{f}_1f_2+\text{h.c.})$. The VMC simulation explicitly reveals how the parton Higgs mass $\lambda$ is generated across the SMG transition in \figref{fig: VMC}(a), providing a deeper understanding of the SMG mechanism from the perspective of the parton Higgs theory \cite{You2018Symmetric, Tong2021Comments}. Note that the parameter $\lambda$ is not accessible in other numerical approaches because it is not a gauge-neutral physical observable. This demonstrates the unique advantage of the VMC approach in studying SMG physics. We expect the methodology to apply to more general SMG phenomena in higher dimensions \cite{Catterall:2016nh,Ayyar:2016tg,Ayyar:2016ph,Schaich2018Phases,Catterall2018Topology,Tong2021Comments,Catterall2209.03828} or Fermi liquids \cite{Lu2210.16304,Zhai2009A0905.1711}, which could have broader implications for the lattice regularization of chiral fermions \cite{Wang2013Non-Perturbative, DeMarco2017A-Novel, Wang2018A-Non-Perturbative, Wang2019Solution, Kikukawa2019Why-is-the-mission, Razamat2021Gapped, Butt2021Anomalies, Zeng2022Symmetric}, the strong-CP problem \cite{Wang2207.14813}, and the pseudo-gap phenomenon \cite{Franzcond-mat/9805401,Kwoncond-mat/9809225,Kwoncond-mat/0006290,Franzcond-mat/0012445,Curtycond-mat/0401124,Zhang2001.09159,Zhang2006.01140,Wang2212.05737}.

\begin{acknowledgments}
We acknowledge the helpful discussions with Tarun Grover, Juven Wang, Da-Chuan Lu, and Meng Zeng. We thank Peiyuan Wang for independently checking our calculations in \appref{app: MF}. The National Science Foundation supported this research under Grant DMR-2238360.
\end{acknowledgments}

\bibliographystyle{apsrev4-2}
\bibliography{ref}

\onecolumngrid
\newpage
\appendix
\setcounter{secnumdepth}{2}
\section{Symmetry Analysis}\label{app: sym}
The bilayer honeycomb lattice model
\begin{equation}\label{eq: H app}
    H=-t\sum_{\langle ij \rangle,l,\sigma}(c^{\dagger}_{il\sigma}c_{jl\sigma}+\text{h.c.})+J\sum_{i}\vect{S}_{i1}\cdot\vect {S}_{i2},
\end{equation}
has an $\U(1)_{1}\times \U(1)_{2}\times \SU(2)\times \Z_{2}^{S}$ internal symmetry together with the honeycomb lattice symmetry. Each symmetry and its corresponding transformation to the fermionic operator is listed in the \tabref{table:1}.
\begin{table}[ht]
\begin{tabular}{ | c | c | c | } 
 \hline
 \multicolumn{3}{|c|}{Internal symmetries} \\
 \hline
 conservation & symmetry group & transformation \\
 \hline
 \makecell[c]{intra-layer\\charge conservation} &$\U(1)_{1}\times \U(1)_{2}$ & $c_{il}\rightarrow e^{-\ii\theta_{l}}c_{il}$  \\
 \hline
 \makecell[c]{inter-layer\\spin conservation} & $\SU(2)$ &$c_{il}\rightarrow e^{-\frac{\ii}{2} \boldsymbol\theta \cdot \boldsymbol\sigma}c_{il}$   \\
 \hline
 \makecell[c]{sub-lattice\\charge conjugation} & $\Z_{2}^{S}$ & \makecell[c]{$c_{il}\rightarrow (-)^{i}c_{il}$\\$\ii\rightarrow -\ii$} \\
 \hline
\end{tabular}
\caption{Internal symmetry of the bilayer honeycomb model.}
\label{table:1}
\end{table}
where $c_{il}=[c_{il\uparrow}, c_{il\downarrow}]^\intercal$. The sub-lattice charge-conjugation symmetry $\Z_{2}^{S}$ let the fermion modes on one sub-lattice pick up a minus sign while keeping the fermion modes on the other sub-lattice unchanged, then take the complex conjugate of both the fermion mode and the imaginary unit $\ii$. Other than the internal symmetries listed above, the system also has honeycomb lattice symmetries such as translation, rotation, reflection etc. Most importantly, the $\Z_{2}^{S}$ and translation symmetry together rule out the possibility of adding any fermionic bilinear terms to gap out the Dirac fermions. Therefore the gapless Dirac semimetal phase is protected by the symmetries at the free-fermion level.

This statement can be proven more explicitly as follows. When $J=0$, the free hopping Hamiltonian on honeycomb lattice gives rise to the graphene band structure that produces $2\times 2\times 2=8$ (two layers, two spins and two valleys) gapless Dirac fermions $\psi_{Ql\sigma}$ at low energy. They can be described by the low-energy effective
field theory Lagrangian
\begin{equation}
    \mathcal{L}=\sum_{Ql\sigma}\Bar{\psi}_{Ql\sigma}\gamma^{\mu}\partial_{\mu}\psi_{Ql\sigma}
\end{equation}
where $Q=K, K'$ labels the two valleys from fermion doubling and $\gamma^{\mu}=(\sigma^{2}, \sigma^{1}, \sigma^{3})$, $\Bar{\psi}_{Ql\sigma}=\psi^{\dagger}_{Ql\sigma}\gamma^{0}$.

Following the existing literature, it is convenient to use the Majorana basis in Hamiltonian formulation, such that eight complex Dirac fermions can be described by an effective Hamiltonian with matrices of size $32\times 32$
\begin{equation}
\begin{split}
    H&=\int\dd^{2} \bm{x}\chi^{\intercal}h_{\times 32}\chi,\\
    h_{\times 32}&=\sum_{a=1}^{2}\ii\partial_{a}\alpha^{a}+\sum_{b}^{5}m_{b}\beta^{b}
\end{split}
\end{equation}
where $\alpha^{a}$ are symmetric matrices, $\beta^{b}$ are anti-symmetric matrices and all the $\alpha^{a}, \beta^{b}$ anti-commute with each other. Here the $\chi^{\intercal}m_{b}\beta^{b}\chi$ terms represent all the possible bilinear mass terms that can be added to the effective theory. The strategy is to list all possible mass terms and then test if they are invariant under the required symmetries. The complete Hamiltonian with all possible bilinear mass terms is
\begin{equation}
\begin{split}
    h_{\times 32}&=\ii\partial_{1}\sigma^{30000}+\ii\partial_{2}\sigma^{13000}+m_{1}\sigma^{20000}\\
    &+m_{2}\sigma^{11120}+m_{3}\sigma^{11200}+m_{4}\sigma^{11320}+m_{5}\sigma^{12000}
\end{split}
\end{equation}
where $\sigma^{ab\cdots}=\sigma^{a}\otimes \sigma^{b}\otimes\cdots$ are the Pauli matrices.

Among the $\U(1)_{1}\times \U(1)_{2}\times \SU(2)\times \Z_{2}^{S}$ symmetries and lattice symmetries we need to concern, it turns out just the translation symmetry plus $\Z_{2}^{S}$ can already rule out all the bilinear mass terms. Following the existing literature, the translation symmetry and $\Z_{2}^{S}$ can be combined to create an emergent anti-unitary symmetry $\Z_{4}^{TF}$ at low-energy: $\psi_{Ql\sigma}\rightarrow\ii\gamma^{0}\psi^{\dagger}_{Ql\sigma}, \ii\rightarrow-\ii$. Translating this symmetry transformation into Majorana basis becomes

\begin{equation}
    \begin{split}
        \Z_{4}^{TF}: \psi_{Ql\sigma}=\begin{pmatrix}
            \psi_{Ql\sigma 1}\\
            \psi_{Ql\sigma 2}
        \end{pmatrix}&\rightarrow\begin{pmatrix}
            -\psi_{Ql\sigma 2}^{\dagger}\\
            \psi_{Ql\sigma 1}^{\dagger}
            \end{pmatrix}, \ii\rightarrow-\ii\\
        \chi_{Ql\sigma}=\begin{pmatrix}
            \chi_{Ql\sigma 11}\\
            \chi_{Ql\sigma 12}\\
            \chi_{Ql\sigma 21}\\
            \chi_{Ql\sigma 22}
        \end{pmatrix}&\rightarrow
        \begin{pmatrix}
            -\ii\chi_{Ql\sigma 21}\\
            \ii\chi_{Ql\sigma 22}\\
            \ii\chi_{Ql\sigma 11}\\
            -\ii\chi_{Ql\sigma 12}
        \end{pmatrix}, \ii\rightarrow-\ii
    \end{split}
\end{equation}
the symmetry transformation in Majorana basis can be written as $\Z_{4}^{TF}: \chi_{Ql\sigma}\rightarrow M\cdot\chi_{Ql\sigma}, \ii\rightarrow-\ii$, where $M=\sigma^{23000}$. Now we can verify the dynamical part of the Hamiltonian is invariant under $M^{\dagger}\cdot \ii\sigma^{a}\cdot M, \ii\rightarrow-\ii, (a=30000, 13000)$ but fails after acting on every bilinear mass terms $M^{\dagger}\cdot\sigma^{b}\cdot M, \ii\rightarrow-\ii, (b=20000, 11120, 11200, 11320, 12000)$.

However, the system can still have an on-site Heisenberg interaction while preserving the internal symmetries, results in a symmetric gap state at strong coupling limit. Therefore, one can expect a SMG transition between both sides. Since there lacks a symmetry-breaking order parameter to distinguish the phases on both sides, such exotic transition is beyond the Landau-Ginzburg-Wilson paradigm and is interesting to study.

\section{Variational Monte Carlo Algorithm}\label{app: VMC}

\subsection{Variational ansatz}
The VMC simulation is based on a mean-field ansatz
\begin{equation}
    H_{\text{MF}}=-t\sum_{\langle ij \rangle l \sigma}c^{\dagger}_{il\sigma}c_{jl\sigma}+\lambda\sum_{i,\sigma}(-)^ic^{\dagger}_{i1\sigma}c_{i2\sigma}+\text{h.c.}.
\end{equation}
The mean-field Hamiltonian $H_{\text{MF}}$ can be solved exactly. Letting $|\Psi_{\text{MF}}[\lambda]\rangle$ be the ground-state of $H_{\text{MF}}$, we assume that the true many-body ground state of the original Hamiltonian \eqnref{eq: H app} can be modeled by $|\Psi_{\text{MF}}[\lambda]\rangle$ followed by a soft projection.
\begin{equation}
    \begin{split}
        |\Psi[\lambda,V]\rangle=&\mathcal{P}[V]|\Psi_{\text{MF}}[\lambda]\rangle\\
        \mathcal{P}[V]&=\mathcal{P}_\text{global}\exp\Big({-\sum_{i}V(q_{i})}\Big)
    \end{split}
\end{equation}
where $q_{i}=(q_{i}^{+},q_{i}^{-})$ is the local charge vector specified by on-site state configurations, defined by
\begin{equation}
q_{i}^{\pm}:=\sum_{l\sigma}(\pm)^{l}c^{\dagger}_{il\sigma}c_{il\sigma}.
\end{equation}
In general, to model the true ground state, $V(q_{i})$ is parameterized by a neural network (a multi-layer feed-forward model) and then optimized by deep learning. In our case, since the Hamiltonian is invariant under charge $\U(1)^{+}$ and layer $\U(1)^{-}$ symmetries, 
\begin{equation}
    \begin{split}
        &Q^{\pm} :=\sum_{i}q_{i}^{\pm} = \sum_{il\sigma}(\pm)^{l}c^{\dagger}_{il\sigma}c_{il\sigma},\\
        &[H,Q^{\pm}]=0.
    \end{split}
\end{equation}
The many-body ground state must be in the global charge neutral $Q^{\pm}=0$ subspace, so a global projection operator $\mathcal{P}_\text{global}$ is introduced to project to the $Q^{\pm}=0$ subspace.

\subsection{Sampling through Markov chain}
Following the existing literature, the mean-field wave function can be computed by an unitary transformation of $H_{\text{MF}}$. Denote the diagonal form of mean-field Hamiltonian as $H_{\text{MF}}=\sum_{\mu}\epsilon_{\mu}\gamma^{\dagger}_{\mu}\gamma_{\mu}$ and eigenstates $|\phi_{\mu}\rangle=\gamma^{\dagger}_{\mu}|0\rangle$, the explicit form of the unitary matrix of system with $L$ sites and four flavours ($\sigma=\uparrow,\downarrow l=1, 2$) is

\begin{equation}
    U=\begin{pmatrix}
\langle1|\phi_{1}\rangle & \dots & \langle1|\phi_{4L}\rangle\\
\vdots & \ddots & \vdots\\
\langle4L|\phi_{1}\rangle & \dots &\langle4L|\phi_{4L}\rangle
\end{pmatrix}.
\end{equation}
The wave-function amplitude can be computed by a Slater determinant
\begin{equation}
\begin{split}
    &\langle x|\Phi_{\text{MF}}\rangle=\langle x_{1},\dots,x_{N_{e}}|\gamma^{\dagger}_{1},\dots,\gamma^{\dagger}_{N_{e}}|0\rangle\\
    &=\det\begin{pmatrix}
\langle x_{1}|\phi_{1}\rangle & \dots & \langle x_{1}|\phi_{N_{e}}\rangle\\
\vdots & \ddots & \vdots\\
\langle x_{N_{e}}|\phi_{1}\rangle & \dots &\langle x_{N_{e}}|\phi_{N_{e}}\rangle
\end{pmatrix}=\det\mathbf{D}.
\end{split}
\end{equation}
where $N_{e}$ is the electron number at half filling in our model. Introducing the auxiliary matrix $\mathbf{M}$, the wave-function ratio can be obtained as
\begin{equation}
    \begin{split}
        \frac{\langle x'|\Phi_{\text{MF}}\rangle}{\langle x|\Phi_{\text{MF}}\rangle}&=\frac{\det\mathbf{D'}}{\det\mathbf{D}}=\sum_{m}D^{-1}_{m\beta}M_{lm}=W_{l\beta},\\
        M&=\begin{pmatrix}
\langle 1|\phi_{1}\rangle & \dots & \langle 1|\phi_{N_{e}}\rangle\\
\vdots & \ddots & \vdots\\
\langle 4L|\phi_{1}\rangle & \dots &\langle 4L|\phi_{N_{e}}\rangle
\end{pmatrix},\\
\mathbf{W}&=\mathbf{M\cdot D^{-1}}.
    \end{split}
\end{equation}
where the configuration $|x'\rangle$ is achieved by moving the $x_{\beta}$ electron on to site $l$. The key point that reduces the computational complexity is that the $\mathbf{W}$ matrix can be updated base on what we already have.
\begin{equation}
    W'_{h\alpha}=W_{h\alpha}-\frac{W_{h\beta}}{W_{l\beta}}(W_{l\alpha}-\delta_{\alpha\beta}).
\end{equation}
Using this update rule, the wave-function amplitude can be computed with least effort along sampling. The samples are collected through Markov chain, such that the algorithm is proceeded in the following way:
\begin{enumerate}
\item Staring from a configuration $|x\rangle$, a new configuration is generated by transition probability $T(x'|x)$.
\item Compute probability ratio $A(x'|x)=|\frac{p(x')}{p(x)}|$, if $A(x'|x)>1$, accept new configuration; if $A(x'|x)<1$, accept new configuration with probability of $A(x'|x)$.
\item The sampling process is then iterated till equilibrium.
\end{enumerate}
After collecting the configurations along sampling, the physical quantity $O(x)$ and probability $p(x)$ of each configuration can be computed to obtain expectation values and the objective function. Finally, the parameters of the model ($\lambda$ in the mean-field ansatz and several other parameters in the potential function $V$) will be updated by backward propagation through the objective function.
\subsection{Compute expectation value and the objective function}
Let ${|x\rangle}_{x\in X}$ be a complete basis of the many-body Hilbert space (with total electric charge and total valley charge neutral). The expectation value of any observable $O$ can be written as
\begin{equation}
    \langle O\rangle=\frac{\langle \Psi|O|\Psi\rangle}{\langle\Psi|\Psi\rangle}=\sum_{x}O(x)p(x),
\end{equation}
with
\begin{equation}
    \begin{split}
    O(x)=\frac{\langle \Psi|O|x\rangle}{\langle \Psi|x\rangle}&=\sum_{x'}\frac{\langle \Psi|x'\rangle}{\langle \Psi|x\rangle}\langle x'|O|x\rangle,\\
    p(x)&=\frac{|\langle \Psi|x\rangle|^2}{\sum_{x}|\langle \Psi|x\rangle|^2}.
    \end{split}
\end{equation}
To sample from $p(x)$, following the Markov chain with the transition probability
\begin{equation}
    p(x'|x)=\frac{p(x')}{p(x)}=\left|\frac{\langle \Psi|x'\rangle}{\langle \Psi|x\rangle}\right|^2.
\end{equation}
The many-body basis $|x\rangle$ is chosen to be the eigen-basis of the projection operator $\mathcal{P}[g]$, such that the amplitude ratio can be evaluated as
\begin{equation}
    \frac{\langle \Psi|x'\rangle}{\langle \Psi|x\rangle}=\frac{\langle \Psi_\text{MF}[u]|x'\rangle}{\langle \Psi_\text{MF}[u]|x\rangle}\frac{\langle x'|\mathcal{P}[E]|x'\rangle}{\langle x|\mathcal{P}[E]|x\rangle}.
\end{equation}
The Mote Carlo expectation value of the physical quantity is approximated by
\begin{equation}
    \langle O\rangle=\sum_{x}O(x)p(x)\sim \frac{1}{N}\sum_{x\in \mathcal{S}}O(x)
\end{equation}
where $\mathcal{S}\in X$ is a multi-subset of all configurations. This approximation lies on the fact that, through Markov chain sampling, the number of visits of a configuration $|x\rangle$ is proportional to its probability $p(x)$. The objective function to miminize is
\begin{equation}
    \begin{split}
    \scL &= \sum_{x}(\langle H\rangle - H(x))\log p_{\theta}(x)\\
    \langle H\rangle&=\sum_{x}H(x)p(x)\sim \frac{1}{N}\sum_{x\in \mathcal{S}}H(x)
    \end{split}
\end{equation}
where $\theta$ denotes the set ofs all parameters of the variational model.

\section{Finite-Size Scaling}\label{app: FSS}

Here we provide examples of extrapolating numerical results of finite spacial system size $(L=5,6,7,8,9,10)$ to one of infinite size. The results of these extrapolations are used to make  \figref{fig: VMC}.

\begin{figure}[ht]
    \includegraphics[width=8cm]{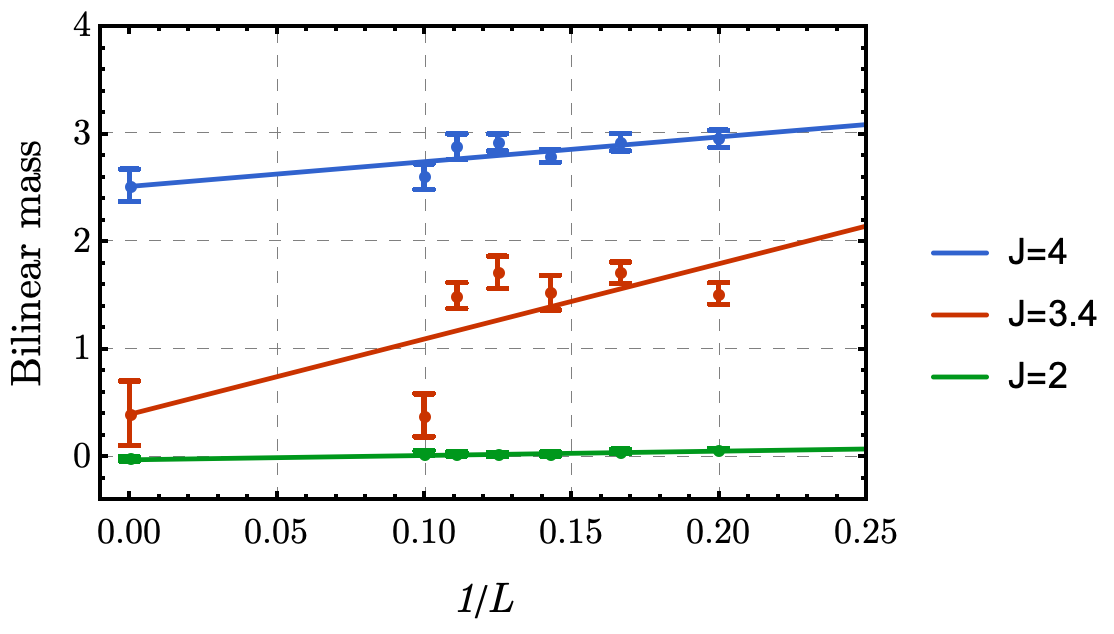}
    \caption{Extrapolation of \figref{fig: VMC}(a)}
\end{figure}
\begin{figure}[ht]
    \includegraphics[width=8cm]{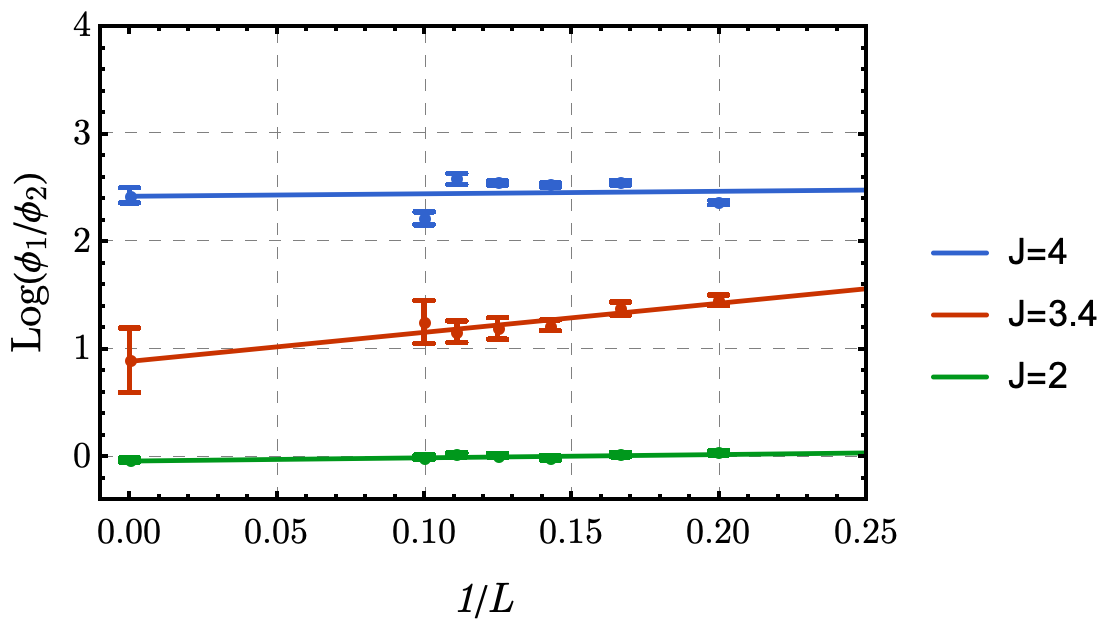}
    \caption{Extrapolation of \figref{fig: VMC}(b)}
\end{figure}
\begin{figure}[ht]
    \includegraphics[width=8cm]{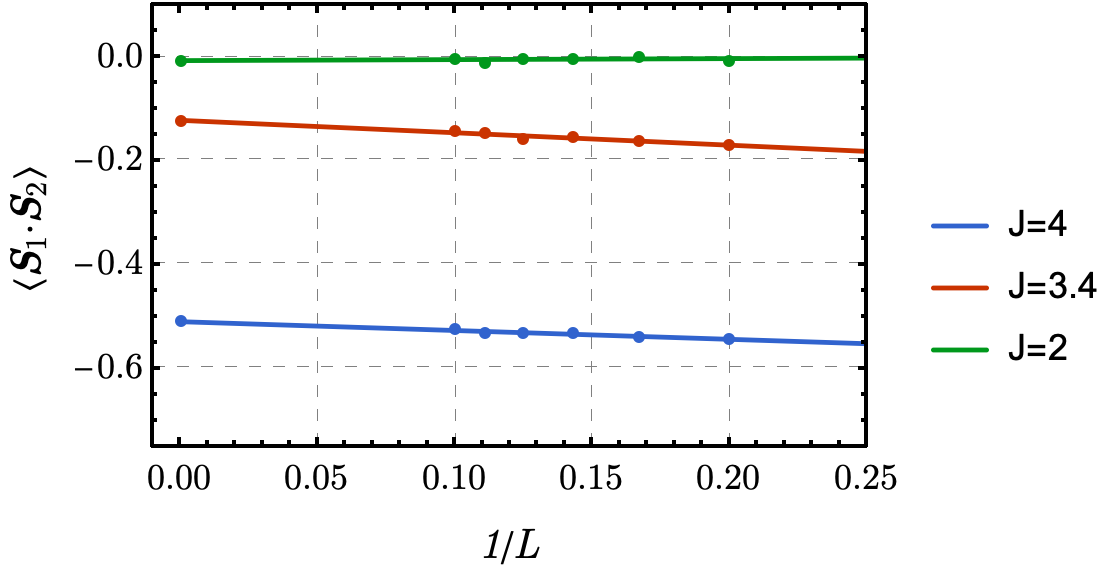}
    \caption{Extrapolation of \figref{fig: VMC}(d)}
\end{figure}

\section{Mean-Field Analaysis} \label{app: MF}
The fermion fractionalization theory can be used to understand the SMG transition. Consider writing an physical fermion mode on each site as a combination of bosonic parton and fermionic parton
\begin{equation}
    c_{il}=\begin{bmatrix}c_{il\uparrow} \\ c_{il\downarrow}\end{bmatrix}=b_{il}\begin{bmatrix}f_{il\uparrow} \\ f_{il\downarrow}\end{bmatrix}=b_{il}f_{il}
\end{equation}
The original Hamiltonian \eqnref{eq: H app} can be rewritten as 
\begin{equation}\label{eq:3}
    H=-t\sum_{\langle ij \rangle l \sigma}b^{\dagger}_{il}b_{jl}f^{\dagger}_{il\sigma}f_{jl\sigma}+h.c.+J\sum_{i}b^{\dagger}_{i1}b_{i1}b^{\dagger}_{i2}b_{i2}\bm{S}^{f}_{i1}\bm{S}^{f}_{i2}
\end{equation}
where the fermionic parton spin has the form $\bm{S}^{f}_{il}=\frac{1}{2}f^{\dagger}_{il\alpha}\sigma^{\alpha\beta}f_{il\beta}$. The partons are redundant descriptions of the original physical fermion, such that the following transformation has no physical effect
\begin{equation}
  \begin{split}
    &b_{il}\rightarrow \e^{-\ii\theta_{il}} b_{il}\\
    &f_{il}\rightarrow \e^{\ii\theta_{il}} f_{il}
  \end{split}
\end{equation}
An emergent gauge group $\Tilde{\U}(1)_{1}\times \Tilde{\U}(1)_{2}$ arises from the fermion fractionalization, the gift of this redundant description is that one can assign the original $\U(1)_{l}$ layer charge onto bosonic parton and write down a bilinear term for fermionic parton without breaking any original symmetry. The charge assignment for physical fermions and bosonic/fermionic partons are specified in \tabref{table:2}. 
\begin{table}[ht]
\begin{tabular}{ c | c c | c c c }
 \ & $\Tilde{\U}(1)_{1}$ & $\Tilde{\U}(1)_{2}$ & $\U(1)_{1}$ & $\U(1)_{2}$ & $\SU(2)$ \\
 \hline
 \ $c_{i1}$ & 0 & 0 & 1 & 0 & 2\\ 
 \ $c_{i2}$ & 0 & 0 & 0 & 1 & 2\\ 
 \ $m_{i}$ & 0 & 0 & 1 & -1 & 1\\ 
 \hline
 \ $b_{i1}$ & -1 & 0 & 1 & 0 & 1\\ 
 \ $b_{i2}$ & 0 & -1 & 0 & 1 & 1\\ 
 \hline
 \ $f_{i1}$ & 1 & 0 & 0 & 0 & 2\\ 
 \ $f_{i2}$ & 0 & 1 & 0 & 0 & 2\\
 \ $M_{i}$ & 1 & -1 & 0 & 0 & 1\\ 
\end{tabular}
\caption{Charge assignment of physical fermion and bosonic/fermionic parton. $m_{i}$ and $M_{i}$ are the bilinear mass of physical fermion and fermionic parton respectively.}
\label{table:2}
\end{table}

We present a mean-field analysis of the fractionalized Hamiltonian \eqnref{eq:3}. The Hamiltonian can be split into bosonic parton part and fermionic parton part under mean-field approximation, such that $H_{MF}=H_{b}+H_{f}$. The Heisenberg interaction $\bm{S}_{i1}\bm{S}_{i2}$ can be further simplified as spin-exchange interaction $f^{\dagger}_{i2}f_{i1} f^{\dagger}_{i1}f_{i2}$ between layers, which plays the same role of spin-spin interaction with only $XY$ components. The bosonic parton Hilbert space is simplified as the Hilbert space of quantum rotor denoted as $b^{\dagger}\sim \e^{\ii\theta_{il}}$.
\begin{table}[ht]
    \centering
    \begin{tabular}{c | c}
        $\langle \e^{\ii\theta_{il}}\rangle$ & $\phi$ \\
        $\langle \e^{-\ii\theta_{i1}}\e^{\ii\theta_{i1}}\e^{-\ii\theta_{i2}}\e^{\ii\theta_{i2}}\rangle$ & $\psi$ \\
        $\langle f^{\dagger}_{il\sigma}f_{jl\sigma}\rangle$ & $u$ \\
        $\langle f^{\dagger}_{i2}f_{i1}\rangle$ & $(-)^{i}2M$ \\
    \end{tabular}
    \caption{Mean-field change of variables}
    \label{tab:b1}
\end{table}

After some change of variables, the mean-field bosonic Hamiltonian becomes on-site and the mean-field fermionic Hamiltonian is quadratic, such that each can be solved self consistently.
\begin{equation}
    \begin{split}
    H_{b}&=\sum_{i} -6 t \phi u \e^{\ii\theta_{il}} +h.c.+ 12t|\phi|^{2}u - J \e^{-\ii\theta_{i1}}\e^{\ii\theta_{i1}}\e^{-\ii\theta_{i2}}\e^{\ii\theta_{i2}} |M|^{2}, \\
    H_{f}&=-t\sum_{\langle ij \rangle l \sigma}|\phi|^{2} f^{\dagger}_{il\sigma}f_{jl\sigma} + h.c.+\frac{J}{2}\psi\sum_{i} (|M|^{2} - (-)^{i}M f^{\dagger}_{i1}f_{i2} - (-)^{i}M^{*}f^{\dagger}_{i2}f_{i1}),
    \end{split}
\end{equation}
where $|\phi|^{2}$ and $|M|^{2}$ play the role of order parameters in bosonic parton and fermionic parton respectively, and the factor of $6=3\times2$ in $H_{b}$ is from the number of bonds on the honeycomb lattice for each site. The ground-state energy per site for each part is
\begin{equation}
    \begin{split}
        E_{b}&=\frac{1}{2}(-J|M|^{2}-24 t u|\phi|^{2}-\sqrt{J^{2}|M|^{4}+576t^{2}u^{2}|\phi|^{2}}),\\
        E_{f}&=\frac{1}{N}\sum^{BZ}_{\bm{k}}J\psi |M|^{2}-\sqrt{J^{2}\psi^{2}|M|^{2}+4t^{2}|\phi|^{4}|f(\bm{k})|^{2}}.
    \end{split}
\end{equation}
In each part, $|\phi|^{2}$ and $|M|^{2}$ are computed by minimizing $E_{b}$ and $E_{f}$ respectively, then the expectation values can be computed self-consistantly as $u=\frac{1}{N}\sum^{BZ}_{\bm{k}}\frac{t|\phi|^{2}|f(\bm{k})|^{2}}{3\sqrt{J^{2}\psi^{2}|M|^{2}+4t^{2}|\phi|^{4}|f(\bm{k})|^{2}}}$ and $\psi=\frac{1}{2}(1+\frac{J|M|^{2}}{\sqrt{J^{2}|M|^{4}+576t^{2}u^{2}|\phi|^{2}}})$.

\begin{figure}[ht]
    \includegraphics[width=8cm]{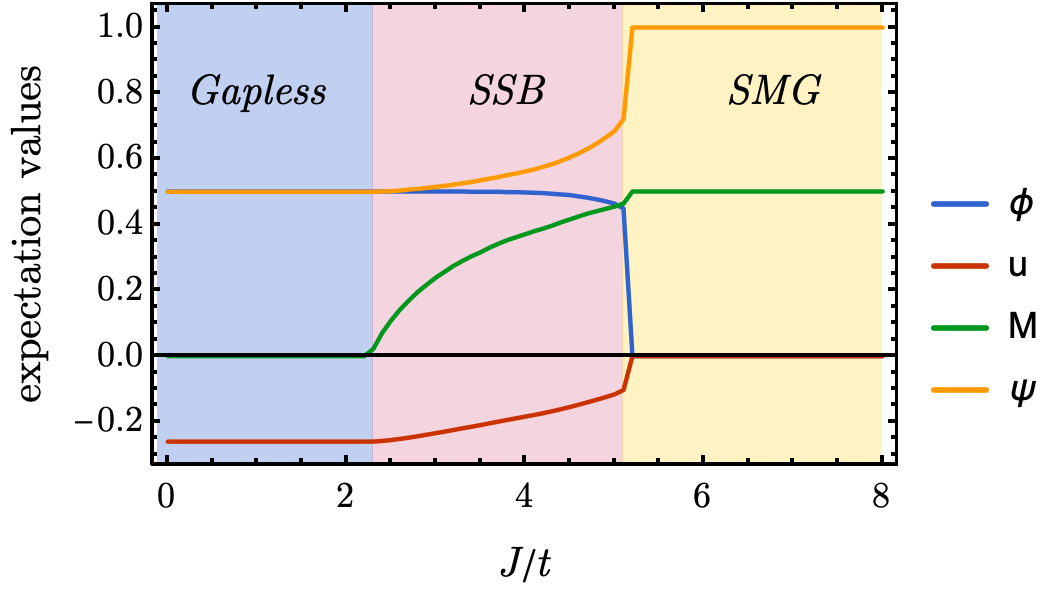}
    \caption{Mean-field result and phase diagram.}
    \label{fig:b1}
\end{figure}

The ground-state optimization is implemented by: 
\begin{enumerate}
\item Input $u$ and $|M|^{2}$ to $E_{b}$ then minimize $E_{b}\big|_{(u,|M|^{2})}$ to get $|\phi|^{2}$, $\psi$.
\item Input $|\phi|^{2}$ and $\psi$ to $E_{f}$ then minimize $E_{f}\big|_{(|\phi|^{2}, \psi)}$ to get $|M|^{2}$, $u$.
\item Iterate till equilibrium.
\end{enumerate}
The result is shown in \figref{fig:b1}.

\begin{itemize}
\item In the weak coupling limit, the fermion site-hopping expectation value recovers the result of the Dirac semimetal $u\sim-0.26$, the inter-layer hopping amplitude $|M|^{2}$ is $0$, and bosonic parton is at super-fluid state $|\phi|^{2}>0$. This corresponds to the gapless Dirac semimetal phase.

\item In the strong coupling limit, the fermionic parton opens up inter-layer hopping order $|M|^{2}>0$, and the bosonic parton is at gap state $|\phi|^{2}=0$. Although the fermionic parton bilinear mass is condensed
\begin{equation}
    H_{\text{M}}=\sum_{i}M_i f^{\dagger}_{i1}f_{i2}+h.c.,
\end{equation}
it does not break any symmetry, but only to Higgs down the gauge group. The resulting state is symmetric and gapped (for both bosonic and fermionic partons), which corresponds to the SMG insulator phase.

\item However, the mean-field theory also predicts an intermediate phase, where $|M|^2>0$ and $|\phi|^2>0$, then $\phi M=m$ will combine into the physical fermion bilinear mass term $m$, which breaks the physical $\U(1)_1\times\U(1)_2$ symmetry. This corresponds to a spontaneous symmetry breaking (SSB) phase, where the system develops inter-layer exciton order. Nevertheless, as we have shown in the main text, the more reliable VMC simulation rules out such an intermediate phase.
\end{itemize}

\begin{figure}[ht]
    \includegraphics[width=7cm]{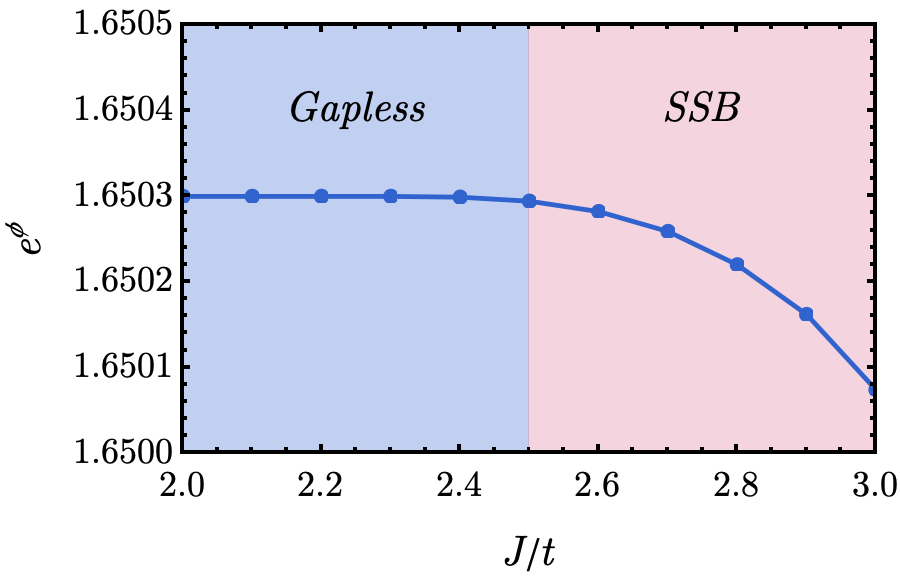}
    \caption{Zoom-in version of the bosonic order parameter}
    \label{fig:b2}
\end{figure}

In the mean-field theory, the gapless-SSB transition is continuous, see \figref{fig:b2}, while the SSB-SMG transition is of first-order (which is obvious from the jump of mean-field parameters).

\end{document}